\documentclass[prl,twocolumn,aps]{revtex4-2}
\usepackage{amsmath}
\usepackage{amssymb}
\usepackage{graphicx}

\begin{document}
\title{Angular Momentum Quantization of a Charge-flux Composite: \\ Quantum Electrodynamic Approach}
\author{Kicheon Kang}
\email{kicheon.kang@gmail.com}
\affiliation{Department of Physics, Chonnam National University, Gwangju 61186, Republic of Korea}

\begin{abstract}
The fractional angular momentum of a two-dimensional charge-flux composite is a well-established phenomenon usually derived from a semiclassical Hamiltonian. However, when the composite is treated as an isolated system in free two-dimensional space, the fundamental rotational and reflection symmetries of the $O(2)$ group demand that its total angular momentum is strictly quantized. We address this conceptual discrepancy by applying a full quantum electrodynamic (QED) approach combined with Noether's theorem. We demonstrate that the interaction between the charge and flux, mediated by the vacuum electromagnetic field, generates an intrinsic interaction angular momentum composed of both field momentum and hidden relativistic momentum. This gauge-invariant interaction angular momentum exactly compensates for the fractional part of the kinetic angular momentum. Consequently, the net angular momentum of the composite strictly follows the integer or half-integer quantization rule. Our formalism clarifies that the conventional fractional spin corresponds to the expectation value of the kinetic angular momentum within the perturbed QED ground state. It also elucidates why the standard classical field angular momentum definition fails to capture this in two dimensions, due to non-vanishing boundary terms.
\end{abstract}

\maketitle

{\em Introduction-}.
The angular momentum ($L$) of any isolated quantum system in free space is quantized as
\begin{equation}
 L = n\hbar \;\;\; \mathrm{or}\;\;\; L = (n+1/2)\hbar ,
 \label{eq:quantized_L}
\end{equation}
where $n$ is an integer. This quantization rule, which is imposed by the rotational and reflection symmetries of space, applies to both two and three spatial dimensions. However, a two-dimensional charge-flux composite does not obey this rule. Its angular momentum is given by $L = n\hbar-q\Phi/2\pi$ (ignoring the spin of each constituent), where $q$ and $\Phi$ are the charge and flux values of each constituent, respectively~\cite{wilczek82}. This composite system is also known as an ``anyon'' and has received significant attention due to its unusual statistical properties, which are generated by fractional spin (for a review, see, e.g., Refs.~\onlinecite{khare05,bartolomei20}). 

This violation of the quantization rule can be explained by the symmetry breaking, specifically the broken reflection symmetry caused by magnetic flux. However, what if we consider the problem from a different perspective, treating the charge and flux as an isolated, composite system in free two dimensions? According to this viewpoint, the composite system would satisfy the quantization rule because the free two-dimensional space has rotational and reflective symmetries. This implies that there must be another contribution to the angular momentum besides $n\hbar- q\Phi/2\pi$. One could attribute the missing angular momentum to the angular momentum of the electromagnetic field, which is defined as (see, for example, Ref.~\onlinecite{griffith12})
\begin{equation}
 \mathbf{L}_f = \epsilon_0 \int  \mathbf{r} \times (\mathbf{E}\times\mathbf{B})\, d^2\mathbf{r} .
 \label{eq:Lf}
\end{equation}
However, this field angular momentum cannot account for the missing contribution. According to the above definition, field momentum is localized at the position of the flux. Furthermore, hidden mechanical momentum~\cite{shockley67,coleman68} exactly compensates for field momentum at the same location. Therefore, the definition of Eq.~\eqref{eq:Lf} does not generate any additional intrinsic angular momentum, and fails to restore the angular momentum quantization.

In this Letter, we show that the quantum electrodynamic (QED) approach combined with Noether's theorem resolves this issue. Instead of using the standard semiclassical description of the interaction between the two bodies, we adopt a complete QED approach. In this approach, the interaction between the two bodies is indirect and is mediated by the vacuum electromagnetic field. Noether's charge associated with rotation gives rise to the conserved angular momentum. Our approach reveals the following. First, the system is an eigenstate of net angular momentum, which consists of two contributions: kinetic and interaction angular momenta. The latter consists of field and hidden relativistic momenta in different locations, which are essentially produced by virtual photons. We find that this interaction part compensates for the fractional value of the kinetic angular momentum and restores the quantization rule of Eq.~\eqref{eq:quantized_L}. The well-known ``fractional spin'' of the composite system corresponds to the expectation value of the kinetic angular momentum of the perturbed QED ground state. Furthermore, the canonical angular momentum is equivalent to the net angular momentum, which is gauge-invariant. This is in sharp contrast with the semiclassical approach.

{\em Angular momentum quantization in two dimensions-}.
We begin with a brief review of the angular momentum quantization in two dimensions. The quantization rule of Eq.~\eqref{eq:quantized_L} is imposed by rotational and reflection symmetries, which are represented by the $O(2)$ group. In two dimensions, the rotation of a quantum system is represented by the unitary operator $D(\phi) = e^{i \hat{L}\phi/\hbar}$, where $\phi$ is the angle of rotation and $\hat{L}$ is the generator of rotation corresponding to angular momentum. After a $2\pi$ rotation, the system returns to its original state. This is represented by the relation $D(2\pi) = e^{2\pi i\alpha}$, where $\alpha$ is an arbitrary real number. This leads to the quantization of the angular momentum as $L=(n+\alpha)\hbar$. 
Reflection symmetry restricts the value of $\alpha$. For the reflection of a state $|\psi\rangle \rightarrow R |\psi\rangle$, the angular momentum transforms as $L\rightarrow R^\dagger L R = -L$. The angular momentum eigenvalue for a reflected state $R|\psi\rangle$ must be $-(n+\alpha)\hbar$, and therefore imposes the condition that either $\alpha=0$ or $\alpha=1/2$.

However, this quantization rule does not seem to apply to a two-dimensional composite of charge ($q$) and magnetic flux ($\Phi$)~(Fig.~1(a)). The system can be described by a single relative coordinate, $\mathbf{x}$, and its Hamiltonian is given by
\begin{equation}
 H = \frac{1}{2\bar{m}} (\mathbf{p}-q\mathbf{A}(\mathbf{x}))^2 ,
\label{eq:H_sc}
\end{equation}
where $\bar{m}$ is the reduced mass and $\mathbf{p} = -i\hbar\nabla$ is the canonical momentum conjugate to the vector $\mathbf{x}$. The angular momentum eigenvalue of this Hamiltonian is $L=n\hbar-q\Phi/2\pi$, with an arbitrary real value of $q\Phi/2\pi$  that does not follow the quantization rule of Eq.~\eqref{eq:quantized_L}. This violation originates from the breaking of the reflection symmetry. In other words, the flux generates a symmetry-breaking vector potential, $\mathbf{A}$. However, when we include the flux as a part of the dynamical system, the composite can be considered as an isolated object in a fully symmetric two-dimensional space (Fig.~1(b)). From this perspective, we expect the quantization rule of Eq.~\eqref{eq:quantized_L} to be restored, as opposed to the arbitrary fractional value of $L$ derived from the semiclassical Hamiltonian (Eq.~\eqref{eq:H_sc}). We will fill this theoretical gap in the following sections.

\begin{figure}[tb]
\centering
\includegraphics[width=8.8cm]{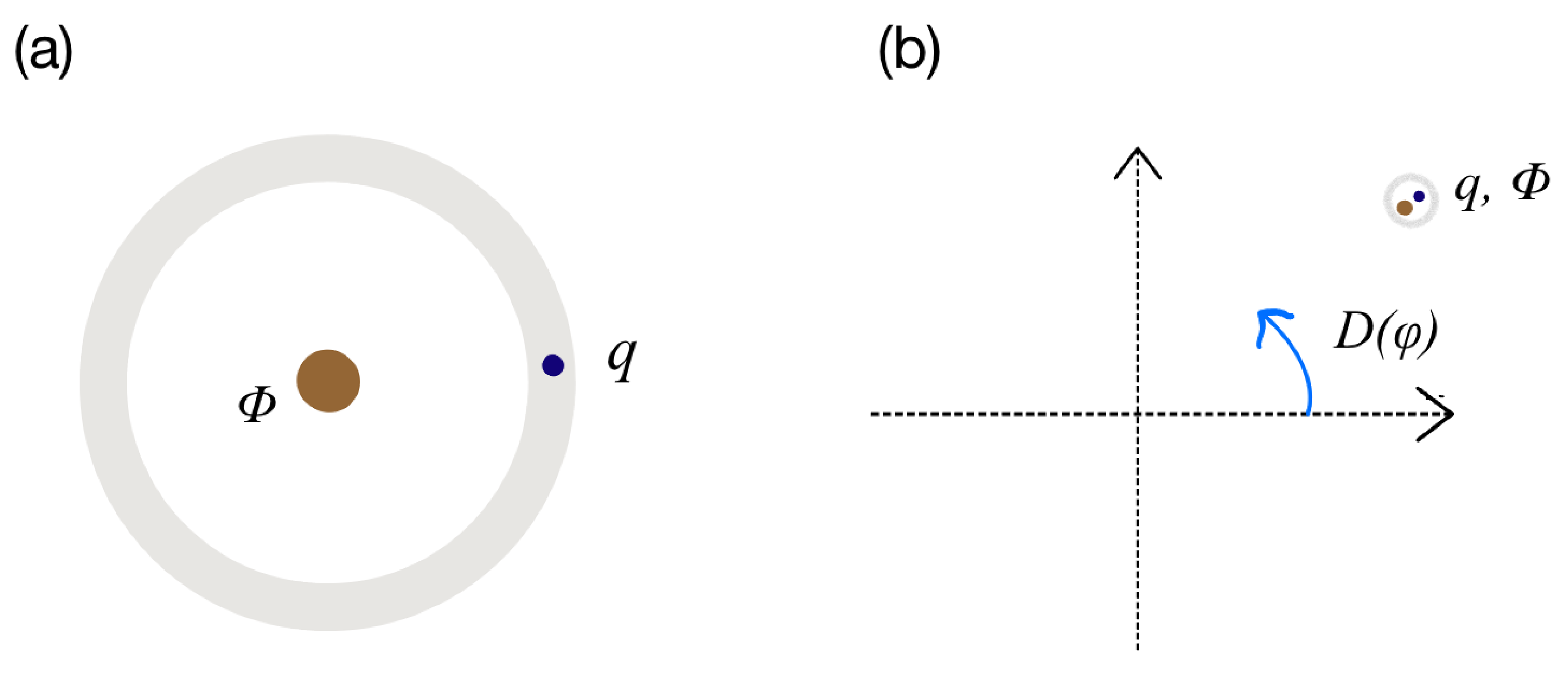} 
\caption{(a) A two-dimensional charge-flux composite consisting of a charge $q$ and a magnetic flux $\Phi$. (b) The composite treated as an isolated quantum system in a fully symmetric free space. 
}
\end{figure}

{\em Attempt to resolve the paradox and its failure in the semiclassical approach-}.
For the composite, $L = n\hbar-q\Phi/2\pi$ corresponds to the kinetic angular momentum. In a system with electromagnetic interaction, there is an additional contribution: the angular momentum of the field, which is given by Eq.~\eqref{eq:Lf}. The field momentum, $\mathbf{P}_f = \epsilon_0\int \mathbf{E}\times\mathbf{B}\, d^2\mathbf{r} = \frac{q\Phi}{2\pi x}\hat{\varphi}$, is localized at the position of the flux, where $\hat{\varphi}$ is the azimuthal unit vector of $\mathbf{x}$. Another important component of the momentum induced by electromagnetic interactions is the hidden relativistic momentum located within the flux. Notably, this momentum exactly cancels out $\mathbf{P}_f$; thus, no additional angular momentum appears. Therefore, the field angular momentum in Eq.~\eqref{eq:Lf} cannot account for the missing angular momentum. In the classical regime, the problem of missing angular momentum is equivalent to Feynman's disk paradox~\cite{feynman10}. This problem can be solved using the angular momentum of the field, given by Eq.~\eqref{eq:Lf}, in three dimensions. In the ``Discussion'' section, we will explain why it cannot be solved using Eq.~\eqref{eq:Lf} in two dimensions due to boundary term scaling.

{\em Quantum electrodynamics and Noether's theorem-}. 
One could argue that the quantization rule in Eq.~\eqref{eq:quantized_L} applies only to canonical angular momentum, not physical angular momentum. This view is mathematically consistent with the formulation of the rule. However, given the profound relationship between symmetry and conservation laws, this perspective is highly unsatisfactory. In our formulation, we explore the possibility of restoring the quantization rule using the quantum electrodynamic (QED) approach. In the QED approach, direct interaction between two bodies at a distance is not presumed. Instead, the interaction is mediated by the vacuum electromagnetic field. This QED perspective offers a deeper understanding of the locality issue concerning the Aharonov-Bohm effect.~\cite{marletto20, kang22}.

The Lagrangian of the composite system can be written as
\begin{subequations}
\begin{eqnarray}
 {\cal L} &=& {\cal L}_0 + {\cal L}_\mathrm{in} ,\\
 {\cal L}_0 &=& \frac{m}{2} \dot{\mathbf{r}}^2 + \frac{M}{2} \dot{\mathbf{R}}^2 +  {\cal L}_\mathrm{em} , \\
 {\cal L}_\mathrm{in} &=& q\dot{\mathbf{r}}  \cdot\mathbf{A}(\mathbf{r}) - qV(\mathbf{r}) \nonumber \\
  &+&  \dot{\mathbf{R}} \cdot \mathbf{a}(\mathbf{R}) + \frac{1}{\mu_0}\vec{\Phi} \cdot \mathbf{B}(\mathbf{R}) ,
\end{eqnarray}
\end{subequations}
where each particle (charge and flux located at $\mathbf{r}$ and $\mathbf{R}$, respectively) interacts locally with the vacuum radiation through vector ($\mathbf{A}$) and scalar ($V$) potentials, as well as electric ($\mathbf{E}$) and magnetic ($\mathbf{B}$) fields, at each position. Here, $\vec{\Phi} = \Phi\hat{z}$ and $\mathbf{a}(\mathbf{R}) \equiv  \epsilon_0\vec{\Phi}\times\mathbf{E}(\mathbf{R})$. The Lagrangian for the radiation is represented by ${\cal L}_\mathrm{em}$. The angular momentum of the composite system can be derived from the Noether charge associated with rotation,
$L= \partial{\cal L}/\partial\dot{\theta} + \partial{\cal L}/\partial\dot{\phi}$, where $\theta$ and $\phi$ represent the angle variables of the vectors $\mathbf{r}$ and $\mathbf{R}$, respectively. 
There is also a contribution from the radiation field. However, this contribution vanishes in the vacuum and is therefore not included here.
We find that
\begin{equation}
 L = L_\mathrm{k} +  L_\mathrm{in} ,
 \label{eq:L}
\end{equation}
where $L_\mathrm{k}$ and $L_\mathrm{in}$ correspond to the kinetic and interaction contributions given by $L_\mathrm{k} = \partial{\cal L}_0/\partial\dot{\theta} + \partial{\cal L}_0/\partial\dot{\phi}$, and $L_\mathrm{in} = \partial{\cal L}_\mathrm{in}/\partial\dot{\theta} + \partial{\cal L}_\mathrm{in}/\partial\dot{\phi}$, respectively.

A standard Legendre transformation, together with quantization, yields a Hamiltonian of the following form:
\begin{eqnarray}
 H &=&  \frac{1}{2m} (\mathbf{p}-q\mathbf{A}(\mathbf{r}))^2 + qV(\mathbf{r}) \nonumber \\
    &+&   \frac{1}{2M} (\mathbf{P}-\mathbf{a}(\mathbf{R}))^2 -  \frac{1}{\mu_0}\vec{\Phi} \cdot \mathbf{B}(\mathbf{R}) + H_\mathrm{em} , 
\label{eq:H}
\end{eqnarray}
where $\mathbf{p}$ and $\mathbf{P}$ are the momentum operators of the charge and the flux, respectively. $H_\mathrm{em}$ represents the radiation modes. 
The radiation field is represented by the standard plane-wave mode expansion of the four-potential (detailed in the Supplemental Material). Our formulation uses the Lorenz gauge in conjunction with the Gupta-Bleuler formalism~\cite{cohen97}, which is essential for accounting for the role of the scalar potential, $V=cA^0$, and enforcing the longitudinal photon condition. In this Hamiltonian formulation, we can write the kinetic and the interaction angular momenta as
\begin{subequations}
\begin{eqnarray}
  L_\mathrm{k} &=& -i\hbar \frac{\partial}{\partial\theta} - q\mathbf{r}\times\mathbf{A} -i\hbar \frac{\partial}{\partial\phi} - \mathbf{R}\times\mathbf{a} , \\
  L_\mathrm{in} &=& q\mathbf{r}\times\mathbf{A} + \mathbf{R}\times\mathbf{a} .
\end{eqnarray}
\end{subequations}
Note that $L_\mathrm{in}$, especially the $q\mathbf{r}\times\mathbf{A}$ term, is not gauge-invariant in the standard semiclassical description; therefore, it is not considered a physical quantity. However, as we will show below, the QED approach eliminates this dependence.
The Hamiltonian (Eq.~\eqref{eq:H}) can be rewritten as $H = H_0 + H_\mathrm{in}$, where $H_0 = \frac{1}{2m} \mathbf{p}^2 + \frac{1}{2M} \mathbf{P}^2 +H_\mathrm{em}$
is the noninteracting part, and 
\begin{equation}
 H_\mathrm{in} = qV(\mathbf{r}) - \frac{q}{2m}(\mathbf{p}\cdot\mathbf{A}+\mathbf{A}\cdot\mathbf{p}) -  \frac{1}{\mu_0}\vec{\Phi} \cdot \mathbf{B} - \frac{1}{2M}(\mathbf{P}\cdot\mathbf{a}+\mathbf{a}\cdot\mathbf{P}) 
\end{equation}
describes the interaction.

Applying standard perturbation theory to the ground state of $H_0$ (see Supplemental Material for details), the dominant contribution comes from the $qV - \frac{1}{\mu_0}\vec{\Phi} \cdot \mathbf{B}$ term. The perturbed ground state is therefore given by $|\psi\rangle = |\phi_0\rangle + (G_0^\dagger + G_b^\dagger) |\phi_0\rangle$, where the operators associated with the scalar ($a_\mathbf{k}^{0\dagger}$) and transverse field ($a_\mathbf{k}^{2\dagger}$) exchanges are given by $G_0^\dagger = -\frac{q c}{\hbar} \sum_\mathbf{k} \frac{\alpha_\mathbf{k}}{\omega} u_\mathbf{k}^*(\mathbf{r}) a_\mathbf{k}^{0\dagger}$ and $G_b^\dagger =  -i \frac{\Phi}{\hbar\mu_0 c} \sum_\mathbf{k} \alpha_\mathbf{k} u_\mathbf{k}^*(\mathbf{R}) a_\mathbf{k}^{2\dagger}$, respectively.

Notably, $|\psi\rangle$ is not an eigenstate of the kinetic angular momentum $L_\mathrm{k}$. This contrasts with the semiclassical approach using the Hamiltonian in Eq.~\eqref{eq:H_sc}. Instead, we calculate the expectation values of $L_\mathrm{k}$ and $L_\mathrm{in}$. We find that $\langle\mathbf{r}\times\mathbf{A}\rangle = \frac{\Phi}{2\pi} \langle\phi_0| \frac{\mathbf{r}\times\hat{\varphi} }{|\mathbf{r}-\mathbf{R}|} |\phi_0 \rangle$ and $\langle\mathbf{R}\times\mathbf{a}\rangle = -\frac{q\Phi}{2\pi} \langle\phi_0| \frac{\mathbf{R}\times\hat{\varphi} }{|\mathbf{r}-\mathbf{R}|} |\phi_0 \rangle$, which leads to our main result:
\begin{subequations}
\label{eq:L-result}
\begin{eqnarray}
  \langle L_\mathrm{k} \rangle &=& n\hbar - \frac{q\Phi}{2\pi} , \\
  \langle L_\mathrm{in} \rangle &=&  \frac{q\Phi}{2\pi} .
\end{eqnarray}
The ground state is an eigenstate of the net angular momentum $L$, that is,
\begin{equation}
  L|\psi\rangle = n\hbar|\psi\rangle .
\end{equation}
\end{subequations}

Thus, the ``fractional spin'' obtained from the standard theory corresponds only to the expectation value of the kinetic angular momentum ($\langle L_\mathrm{k} \rangle$) in our approach. However, the standard theory only accounts for part of the total angular momentum; the rest is generated by interactions ($\langle L_\mathrm{in} \rangle$). Interaction angular momentum consists of field momentum located at the charge and hidden mechanical momentum at the flux. Consequently, the quantization rule is fully restored for the net angular momentum ($L$), and the true ground state is an eigenstate of $L$.

{\em A quantum analog of Feynman's disk paradox-}.
To further elucidate the physical implications of this quantization, it is instructive to consider a two-dimensional quantum analog of Feynman's disk paradox~\cite{feynman10}. Let us configure the charge-flux composite such that the magnetic flux $\Phi$ is generated by two infinitesimally close, counter-rotating loops carrying opposite charges~(Fig.~2). According to the standard semiclassical description, the external charge $q$ is in an eigenstate of the kinetic angular momentum with the fractional eigenvalue $L_\mathrm{k} = n\hbar - q\Phi/2\pi$.

Now, suppose a weak internal dissipative interaction (such as radiative damping) slows the counter-rotating loops. This eventually brings them to a complete rest, leading to an adiabatic switching-off of the magnetic flux. As the flux decreases to zero, the resulting induced azimuthal electric field imparts an additional angular momentum of exactly $+q\Phi/2\pi$ to the charge $q$ via Faraday's law of induction. At the same time, whatever initial mechanical angular momentum possessed by the loops is transferred entirely to the surrounding vacuum via the emission of photons. Since the angular momentum of the electromagnetic field is quantized, this radiative transfer of angular momentum must strictly occur in discrete, integer units of $m\hbar$.

Consequently, in the final state where the flux is completely extinguished and the loops are stationary, the net angular momentum of the entire isolated system-comprising the charge, the loops, and the vacuum radiation field must be an exact integer multiple of $\hbar$, $(n+m)\hbar$. By the fundamental conservation of total angular momentum, the net angular momentum of the initial configuration must also have been strictly quantized in units of $\hbar$. This thought experiment clearly shows that the arbitrary fractional value derived from the semiclassical kinetic term is incomplete because the system has an exact compensating interaction angular momentum.

\begin{figure}[tb]
\centering
\includegraphics[width=5cm]{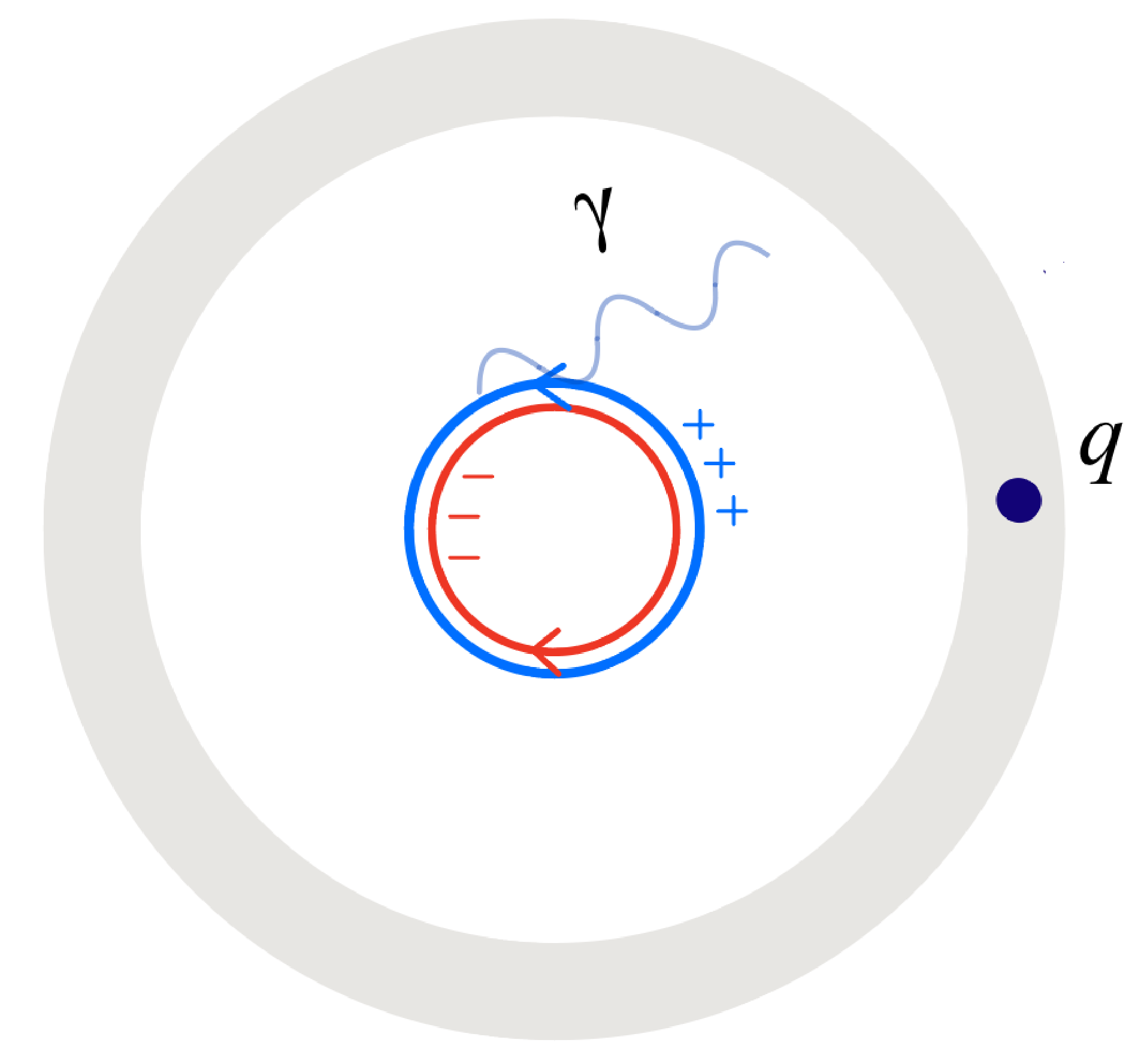} 
\caption{A two-dimensional quantum analog of Feynman's disk paradox.  Two infinitesimally close, counter-rotating loops carrying opposite charges generate the central magnetic flux. A weak internal dissipative interaction, such as radiative damping, slows the loops to a complete rest. This allows the transfer of their mechanical angular momentum to the surrounding vacuum via the emission of photons ($\gamma$).
 Meanwhile, the adiabatic switching off of the flux induces an azimuthal electric field that imparts an additional angular momentum of exactly $+q\Phi/2\pi$ to the external charge $q$ via Faraday's law of induction.}
\end{figure}

{\em Discussion-}.
It is crucial to understand why the standard field angular momentum integral $\mathbf{L}_{em}^f$ defined in Eq.~\eqref{eq:Lf}, fails to resolve the missing angular momentum in two dimensions. The interaction angular momentum derived directly from Noether's theorem takes the form $\mathbf{L}_{em}^A = \int \rho \mathbf{r}\times\mathbf{A}\, d^2\mathbf{r}$. Working within the Coulomb gauge ($\nabla \cdot \mathbf{A} = 0$), the relationship between these two definitions can be expressed in a compact form as $\mathbf{L}_{em}^f = \mathbf{L}_{em}^A + \epsilon_0 \int \nabla \cdot \mathbf{Q}\, d\tau$. Here, the tensor $\mathbf{Q}$ is defined by its components $Q_{ij} = \epsilon_i^{lk} r_l \left[ (\mathbf{E}\cdot\mathbf{A})\delta_{kj} - (E_j A_k + E_k A_j) \right]$. 

In three dimensions, the fields decay sufficiently fast such that the terms composing $Q_{ij}$ (scaling as $r E A$) scale as $\lesssim 1/r^3$. Consequently, the boundary integral $\int \nabla \cdot \mathbf{Q}\, d\tau = \oint \mathbf{Q} \cdot d\mathbf{a}$ vanishes as $r \to \infty$, rendering $\mathbf{L}_{em}^A$ and $\mathbf{L}_{em}^f$ mathematically equivalent (see e.g., Ref.~\onlinecite{aguirregabiria81}). However, the situation is fundamentally different for a charge and flux composite in two dimensions. Because the vector potential $\mathbf{A}$ and the electric field $\mathbf{E}$ both scale as $1/r$, the boundary tensor components scale as $Q_{ij} \sim rEA \propto 1/r$. Consequently, when taking the limit $r \to \infty$, the boundary integral $\oint \mathbf{Q} \cdot d\mathbf{a}$ does not vanish. This non-vanishing boundary term explicitly breaks the equivalence between the two definitions in 2D. Our results suggest that the physically relevant quantity to use is $\mathbf{L}_{em}^A$, a rigorously derived vector from Noether's theorem. This explains why the standard cross-product field momentum integral cannot accurately account for the expected fractional angular momentum in a two-dimensional (2D) configuration.

Another important aspect of our approach is the gauge invariance of the interaction angular momentum, $\langle L_\mathrm{in} \rangle$. In the semiclassical approach, the term $\mathbf{r}\times\mathbf{A}$ is inherently gauge-dependent, presenting a significant theoretical hurdle. In the QED framework, however, the expectation value $\langle\mathbf{r}\times\mathbf{A}\rangle$ exclusively involves the exchange of transverse photons, ensuring that no gauge ambiguity affects the intrinsic interaction. 
Note that the longitudinal component of $\mathbf{A}$ contributes to the interaction Hamiltonian of $-(q/2m)(\mathbf{p}\cdot\mathbf{A}+\mathbf{A}\cdot\mathbf{p})$, which yields an additional term proportional to $\nabla\cdot\mathbf{A}$. However, this term only contributes to the self-interaction of the charge and not to the mutual interaction between two bodies.
Furthermore,  this contribution disappears in our QED approach.
Meanwhile, the hidden momentum term ($\langle\mathbf{R}\times\mathbf{a}\rangle$), which is mediated by scalar photons, involves only the gauge-invariant quantity, $\mathbf{a}$. Therefore, our QED formulation fully resolves the gauge issues inherent in the standard semiclassical description.

{\em Conclusion-}.
In summary, we have demonstrated that the angular momentum quantization in 2D charge-flux composites is restored through a complete QED formulation and Noether's theorem. By acknowledging that the interaction mediated by the vacuum electromagnetic field possesses both field momentum and hidden mechanical momentum, the net angular momentum perfectly recovers its integer $n\hbar$ quantization. Furthermore, our formalism clarifies that the widely accepted ``fractional spin'' does not indicate a breakdown of fundamental spatial symmetries; rather, it corresponds to the expectation value of the kinetic contribution. By establishing the gauge invariance of the interaction angular momentum and demonstrating the failure of the classical field angular momentum, our work provides a complete and rigorous QED view for the physics of anyons.

 \bibliography{references}

\begin{thebibliography}{11}%
\makeatletter
\providecommand \@ifxundefined [1]{%
 \@ifx{#1\undefined}
}%
\providecommand \@ifnum [1]{%
 \ifnum #1\expandafter \@firstoftwo
 \else \expandafter \@secondoftwo
 \fi
}%
\providecommand \@ifx [1]{%
 \ifx #1\expandafter \@firstoftwo
 \else \expandafter \@secondoftwo
 \fi
}%
\providecommand \natexlab [1]{#1}%
\providecommand \enquote  [1]{``#1''}%
\providecommand \bibnamefont  [1]{#1}%
\providecommand \bibfnamefont [1]{#1}%
\providecommand \citenamefont [1]{#1}%
\providecommand \href@noop [0]{\@secondoftwo}%
\providecommand \href [0]{\begingroup \@sanitize@url \@href}%
\providecommand \@href[1]{\@@startlink{#1}\@@href}%
\providecommand \@@href[1]{\endgroup#1\@@endlink}%
\providecommand \@sanitize@url [0]{\catcode `\\12\catcode `\$12\catcode
  `\&12\catcode `\#12\catcode `\^12\catcode `\_12\catcode `\%12\relax}%
\providecommand \@@startlink[1]{}%
\providecommand \@@endlink[0]{}%
\providecommand \url  [0]{\begingroup\@sanitize@url \@url }%
\providecommand \@url [1]{\endgroup\@href {#1}{\urlprefix }}%
\providecommand \urlprefix  [0]{URL }%
\providecommand \Eprint [0]{\href }%
\providecommand \doibase [0]{https://doi.org/}%
\providecommand \selectlanguage [0]{\@gobble}%
\providecommand \bibinfo  [0]{\@secondoftwo}%
\providecommand \bibfield  [0]{\@secondoftwo}%
\providecommand \translation [1]{[#1]}%
\providecommand \BibitemOpen [0]{}%
\providecommand \bibitemStop [0]{}%
\providecommand \bibitemNoStop [0]{.\EOS\space}%
\providecommand \EOS [0]{\spacefactor3000\relax}%
\providecommand \BibitemShut  [1]{\csname bibitem#1\endcsname}%
\let\auto@bib@innerbib\@empty
\bibitem [{\citenamefont {Wilczek}(1982)}]{wilczek82}%
  \BibitemOpen
  \bibfield  {author} {\bibinfo {author} {\bibfnamefont {F.}~\bibnamefont
  {Wilczek}},\ }\bibfield  {title} {\bibinfo {title} {Magnetic flux, angular
  momentum, and statistics},\ }\href@noop {} {\bibfield  {journal} {\bibinfo
  {journal} {Phys. Rev. Lett.}\ }\textbf {\bibinfo {volume} {48}},\ \bibinfo
  {pages} {1144} (\bibinfo {year} {1982})}\BibitemShut {NoStop}%
\bibitem [{\citenamefont {Khare}(2005)}]{khare05}%
  \BibitemOpen
  \bibfield  {author} {\bibinfo {author} {\bibfnamefont {A.}~\bibnamefont
  {Khare}},\ }\href@noop {} {\emph {\bibinfo {title} {Fractional statistics and
  quantum theory}}}\ (\bibinfo  {publisher} {World Scientific},\ \bibinfo
  {year} {2005})\BibitemShut {NoStop}%
\bibitem [{\citenamefont {Bartolomei}\ \emph {et~al.}(2020)\citenamefont
  {Bartolomei}, \citenamefont {Kumar}, \citenamefont {Bisognin}, \citenamefont
  {Marguerite}, \citenamefont {Berroir}, \citenamefont {Bocquillon},
  \citenamefont {Placais}, \citenamefont {Cavanna}, \citenamefont {Dong},
  \citenamefont {Gennser} \emph {et~al.}}]{bartolomei20}%
  \BibitemOpen
  \bibfield  {author} {\bibinfo {author} {\bibfnamefont {H.}~\bibnamefont
  {Bartolomei}}, \bibinfo {author} {\bibfnamefont {M.}~\bibnamefont {Kumar}},
  \bibinfo {author} {\bibfnamefont {R.}~\bibnamefont {Bisognin}}, \bibinfo
  {author} {\bibfnamefont {A.}~\bibnamefont {Marguerite}}, \bibinfo {author}
  {\bibfnamefont {J.-M.}\ \bibnamefont {Berroir}}, \bibinfo {author}
  {\bibfnamefont {E.}~\bibnamefont {Bocquillon}}, \bibinfo {author}
  {\bibfnamefont {B.}~\bibnamefont {Placais}}, \bibinfo {author} {\bibfnamefont
  {A.}~\bibnamefont {Cavanna}}, \bibinfo {author} {\bibfnamefont
  {Q.}~\bibnamefont {Dong}}, \bibinfo {author} {\bibfnamefont {U.}~\bibnamefont
  {Gennser}}, \emph {et~al.},\ }\bibfield  {title} {\bibinfo {title}
  {Fractional statistics in anyon collisions},\ }\href@noop {} {\bibfield
  {journal} {\bibinfo  {journal} {Science}\ }\textbf {\bibinfo {volume}
  {368}},\ \bibinfo {pages} {173} (\bibinfo {year} {2020})}\BibitemShut
  {NoStop}%
\bibitem [{\citenamefont {Griffith}(2012)}]{griffith12}%
  \BibitemOpen
  \bibfield  {author} {\bibinfo {author} {\bibfnamefont {D.~J.}\ \bibnamefont
  {Griffith}},\ }\href@noop {} {\bibinfo {title} {Introduction to
  electrodynamics 4th edition, sec. 8.2}} (\bibinfo {year} {2012})\BibitemShut
  {NoStop}%
\bibitem [{\citenamefont {Shockley}\ and\ \citenamefont
  {James}(1967)}]{shockley67}%
  \BibitemOpen
  \bibfield  {author} {\bibinfo {author} {\bibfnamefont {W.}~\bibnamefont
  {Shockley}}\ and\ \bibinfo {author} {\bibfnamefont {R.}~\bibnamefont
  {James}},\ }\bibfield  {title} {\bibinfo {title} {Try simplest cases
  discovery of hidden momentum forces on magnetic currents},\ }\href@noop {}
  {\bibfield  {journal} {\bibinfo  {journal} {Phys. Rev. Lett.}\ }\textbf
  {\bibinfo {volume} {18}},\ \bibinfo {pages} {876} (\bibinfo {year}
  {1967})}\BibitemShut {NoStop}%
\bibitem [{\citenamefont {Coleman}\ and\ \citenamefont
  {Van~Vleck}(1968)}]{coleman68}%
  \BibitemOpen
  \bibfield  {author} {\bibinfo {author} {\bibfnamefont {S.}~\bibnamefont
  {Coleman}}\ and\ \bibinfo {author} {\bibfnamefont {J.}~\bibnamefont
  {Van~Vleck}},\ }\bibfield  {title} {\bibinfo {title} {Origin of ``hidden
  momentum forces" on magnets},\ }\href@noop {} {\bibfield  {journal} {\bibinfo
   {journal} {Phys. Rev.}\ }\textbf {\bibinfo {volume} {171}},\ \bibinfo
  {pages} {1370} (\bibinfo {year} {1968})}\BibitemShut {NoStop}%
\bibitem [{\citenamefont {Feynman}\ \emph {et~al.}(2010)\citenamefont
  {Feynman}, \citenamefont {Leighton},\ and\ \citenamefont
  {Sands}}]{feynman10}%
  \BibitemOpen
  \bibfield  {author} {\bibinfo {author} {\bibfnamefont {R.~P.}\ \bibnamefont
  {Feynman}}, \bibinfo {author} {\bibfnamefont {R.~B.}\ \bibnamefont
  {Leighton}},\ and\ \bibinfo {author} {\bibfnamefont {M.}~\bibnamefont
  {Sands}},\ }\href@noop {} {\bibinfo {title} {The feynman lectures on physics
  ii, the new millennium edition, sec.17-4}} (\bibinfo {year}
  {2010})\BibitemShut {NoStop}%
\bibitem [{\citenamefont {Marletto}\ and\ \citenamefont
  {Vedral}(2020)}]{marletto20}%
  \BibitemOpen
  \bibfield  {author} {\bibinfo {author} {\bibfnamefont {C.}~\bibnamefont
  {Marletto}}\ and\ \bibinfo {author} {\bibfnamefont {V.}~\bibnamefont
  {Vedral}},\ }\bibfield  {title} {\bibinfo {title} {Aharonov-bohm phase is
  locally generated like all other quantum phases},\ }\href@noop {} {\bibfield
  {journal} {\bibinfo  {journal} {Phys. Rev. Lett.}\ }\textbf {\bibinfo
  {volume} {125}},\ \bibinfo {pages} {040401} (\bibinfo {year}
  {2020})}\BibitemShut {NoStop}%
\bibitem [{\citenamefont {Kang}(2022)}]{kang22}%
  \BibitemOpen
  \bibfield  {author} {\bibinfo {author} {\bibfnamefont {K.}~\bibnamefont
  {Kang}},\ }\bibfield  {title} {\bibinfo {title} {Gauge invariance of the
  local phase in the aharonov-bohm interference: Quantum electrodynamic
  approach},\ }\href {https://doi.org/10.1209/0295-5075/ac9fee} {\bibfield
  {journal} {\bibinfo  {journal} {Europhys. Lett.}\ }\textbf {\bibinfo {volume}
  {140}},\ \bibinfo {pages} {46001} (\bibinfo {year} {2022})}\BibitemShut
  {NoStop}%
\bibitem [{\citenamefont {Cohen-Tannoudji}\ \emph {et~al.}(1997)\citenamefont
  {Cohen-Tannoudji}, \citenamefont {Dupont-Roc},\ and\ \citenamefont
  {Grynberg}}]{cohen97}%
  \BibitemOpen
  \bibfield  {author} {\bibinfo {author} {\bibfnamefont {C.}~\bibnamefont
  {Cohen-Tannoudji}}, \bibinfo {author} {\bibfnamefont {J.}~\bibnamefont
  {Dupont-Roc}},\ and\ \bibinfo {author} {\bibfnamefont {G.}~\bibnamefont
  {Grynberg}},\ }\href@noop {} {\emph {\bibinfo {title} {Photons and
  Atoms-Introduction to Quantum Electrodynamics}}}\ (\bibinfo {year}
  {1997})\BibitemShut {NoStop}%
\bibitem [{\citenamefont {Aguirregabiria}\ and\ \citenamefont
  {Hern{\'a}ndez}(1981)}]{aguirregabiria81}%
  \BibitemOpen
  \bibfield  {author} {\bibinfo {author} {\bibfnamefont {J.}~\bibnamefont
  {Aguirregabiria}}\ and\ \bibinfo {author} {\bibfnamefont {A.}~\bibnamefont
  {Hern{\'a}ndez}},\ }\bibfield  {title} {\bibinfo {title} {The feynman paradox
  revisited},\ }\href@noop {} {\bibfield  {journal} {\bibinfo  {journal} {Eur.
  J. Phys.}\ }\textbf {\bibinfo {volume} {2}},\ \bibinfo {pages} {168}
  (\bibinfo {year} {1981})}\BibitemShut {NoStop}%
\end{thebibliography}%

\end{document}